\documentclass[a4paper,12pt]{article}
\usepackage{graphics}
\usepackage{amsmath,amssymb}
\def\vf{\varphi}
\def\ve{\varepsilon}
\def\half{\frac{1}{2}}
\def\be{\begin{equation}}
\def\ee{\end{equation}}
\def\bea{\begin{eqnarray}}
\def\eea{\end{eqnarray}}
\def\beax{\begin{eqnarray*}}
\def\eeax{\end{eqnarray*}}
\title
{On Scattering from Localized non-Hermitian Potentials}
\author{H.~F.~Jones\thanks{h.f.jones@imperial.ac.uk}\\
Imperial College London}
\date{}

\begin{document}
\setcounter{footnote}{1}
\renewcommand{\thefootnote}{{\fnsymbol{footnote}}}
\maketitle
\begin{abstract}
We highlight the conceptual issues that arise when one applies the quasi-Hermitian framework
to analyze  scattering from localized non-Hermitian potentials, in particular complex square-wells
or delta-functions. When treated in the framework of conventional quantum mechanics, these potentials
are generally considered as effective theories, in which probability is not conserved because of processes that
have been ignored. However, if they are treated as fundamental theories, the Hilbert-space metric
must be changed. In order for the newly-defined probability to be conserved, it must differ from the standard one,
even at asymptotically large distances from the scattering centre, and the mechanism for this is the non-locality of the new metric,
as we show in detail in the model of a single complex delta function.
However, properties of distant bound-state systems, which do not interact physically with the non-Hermitian scattering potential,
should not be affected. We analyze a model Hamiltonian that supports this contention.
\\


\end{abstract}
\section{Introduction}

Since the seminal paper by Bender and Boettcher\cite{BB} on $PT$-symmetric but non-Hermitian
Hamiltonians, subsequent research has gone through various stages.
First came an exploration of the reality
or otherwise of the spectrum of non-Hermitian generalizations of well-known soluble models (see
Ref.~\cite{GZ} for a systematic approach).
However, it soon became apparent that something additional to the reality of the spectrum was
needed if such models were to be viable as quantum theories with a proper probabilistic interpretation.
This is because the natural metric for $PT$-symmetric models gives the overlap of two wave-functions
$\psi(x)$ and $\vf(x)$ as $\int dx \vf^*(-x) \psi(x)$, rather than the usual $\int dx \vf^*(x) \psi(x)$.
Since the corresponding norm is not positive definite, the theory endowed with this metric does not
represent a physical framework for quantum mechanics. Instead it turns out to be possible to
construct\cite{BBJ, Hendrik, AM-metric}
an alternative, positive-definite metric $\eta$, which is dynamically determined by the particular
Hamiltonian in question (see Eq.~(\ref{psH}) below). It was also shown \cite{AM-h} that $\eta$ provided a
similarity transformation from the non-Hermitian $H$ to an equivalent Hermitian $h$, and this equivalent
Hermitian Hamiltonian was subsequently constructed, frequently in perturbation theory only, in a variety
of models\cite{HFJ-ix3, AM-ix3, -x4}.

These investigations were all concerned with isolated non-Hermitian systems, but more recently some attention
has been given to situations where a non-Hermitian system interacts with the world of Hermitian quantum mechanics.
For example, Ref.~\cite{QB} examined a non-Hermitian analogue of the Stern-Gerlach experiment in which the role of the
intermediate inhomogeneous magnetic field flipping the spin is taken over by an apparatus described by a non-Hermitian Hamiltonian.
This type of set-up has been further elaborated by Assis and Fring\cite{Andreas} and Guenther al.\cite{Uwe}, although the analysis
has been challenged by Martin\cite{Martin} and Mostafazadeh\cite{AM-QB}. In another investigation\cite{coupling}
we show in a number of examples that when a non-Hermitian Hamiltonian with a real spectrum is coupled to a Hermitian system
the resulting combined system also has a real spectrum provided that the coupling constant does not exceed a critical value.

It is in this spirit that we reconsider here the conceptual issues involved in scattering from a local non-Hermitian potential,
first addressed in a somewhat different context by Znojil\cite{Znojil},
taking as specific examples a pair of $PT$-symmetric complex $\delta$ functions, a finite $PT$-symmetric complex square
well\cite{ZA, AM-SqWell, Cannata}, and a single complex $\delta$ function\cite{AM-1DF}.

In the next section we review the elementary calculations that show that for such potentials the
probability, as usually defined, is not conserved. This was not seen as a problem by the authors
of Refs.~\cite{ZA, Cannata}, who presumably took the point of view that these potentials were
to be treated as {\it effective} potentials, as is frequently done in nuclear physics,
the complex nature of the potential representing processes not taken into account, where typically
the particle can be absorbed rather than merely scattered.

However, if one takes these potentials seriously as fundamental interactions, then the whole tenor of
the work summarized above is that one should change the Hilbert space by adopting the appropriate metric.
Of necessity, this metric must differ from the standard one not only in the vicinity of the
non-Hermitian potential, but also at distances remote from it, in order that the newly-defined probability
be conserved.

In Section 3 we reanalyze the case of the single complex delta function in the quasi-Hermitian framework.
We will see how this gives a mathematical resolution of the problem, and it is of some interest to see
how it works in detail. The essential mechanism is the non-locality of the metric.

In the final section we address the conceptual issues arising from these calculations, and in particular
emphasize the special status of the plane waves, which are substantially affected by the change in the metric,
as opposed to bound-state wave functions, where a model calculation suggests that the changes induced by the
introduction of a distant non-Hermitian potential are exponentially suppressed.

\section{Some Simple Complex Scattering Potentials}
\subsection{Two delta functions}
The simplest such potential to consider, designed to be $PT$-symmetric, is a pair of imaginary $\delta$ functions:
\bea
V(x)=i\lambda(\delta(x-a)-\delta(x+a))
\eea
with a wave-function of the form
\bea
\psi=\left\{\begin{array}{ll}
e^{ikx}+C e^{-ikx}&\hspace{1cm} x< -a\\
A e^{ikx}+C e^{-ikx}&\hspace{1cm} x< -a\\
D e^{ikx}&\hspace{1cm} x< -a
\end{array}\right.
\eea
Applying the continuity conditions
\bea
\begin{array}{lcl}
\psi_{+}&=&\psi{-}\\
\psi'_{+}&=& \psi'_{-}\pm i\lambda \psi
\end{array}
\eea
at each of the interfaces, we readily find that
\bea
D&=&\frac{1}{1+2i\alpha^2e^{2ika} \sin{2ka}}\nonumber\\
&&\\
C&=& 2iD \alpha(1-\alpha) \sin{2ka},\nonumber
\eea
where $\alpha\equiv \lambda/(2k)$.\\
Calculating $|D|^2$ and $|C|^2$, which in conventional quantum mechanics
represent the respective probabilities for transmission and reflection we find that
\bea
|D|^2=\frac{1}{1-4\alpha^2(1-\alpha^2)\sin^2{2ka}},
\eea
which is always greater than 1 for $\alpha<1$, as of course is the total probability
\bea
|C|^2+|D|^2=\frac{1+4\alpha^2(1-\alpha)^2\sin^2{2ka}}{1-4\alpha^2(1-\alpha^2)\sin^2{2ka}}.
\eea
\subsection{Imaginary Square Well}
A somewhat more complicated potential, still $PT$-symmetric, is the imaginary square-well potential
\cite{ZA, AM-SqWell, Cannata}
\bea\label{sqwell}
V(x)=\left\{\begin{array}{ll}
0 &\hspace{1cm} |x|> a\\
-i\lambda &\hspace{1cm} -a< x< 0\\
i\lambda&\hspace{1cm} 0<x<a
\end{array}\right. ,
\eea
for which a physical realization, as an effective potential, was given by Ruschhaupt et al. \cite{Ruschhaupt}.
The expressions for the standard transmission and reflection coefficients are now
considerably more involved\cite{ZA, Cannata} but Fig.~1, which is a graph of
the transmission coefficient $T$ for $a=\lambda=1$, shows that it exceeds unity for a wide range of values of $k$.
\begin{figure}[h]
\resizebox{!}{2.5in}{\includegraphics{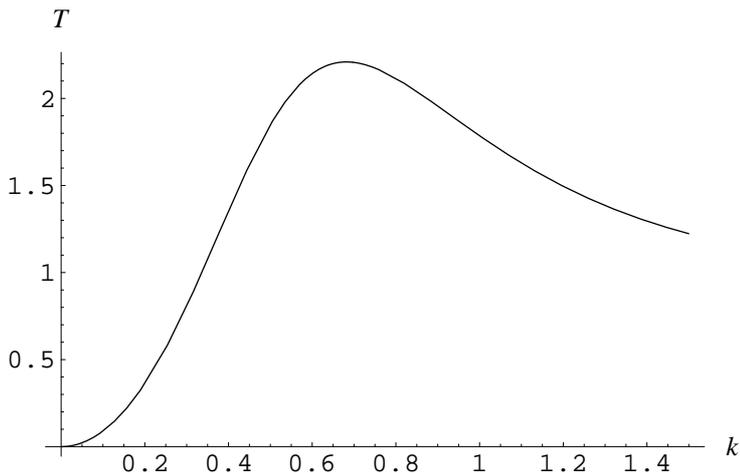}}
\caption{Standard transmission coefficient of Eq.~(\ref{sqwell}) for $a=\lambda=1$}
\end{figure}

\subsection{Single Complex Delta Function}
Here we consider a single complex delta-function potential of the form
\bea\label{1DF}
V=z\delta(x),
\eea
where $z$ is complex.
The wave-function now takes the simple form
\bea\label{psi}
\psi=\left\{\begin{array}{ll}
e^{ikx}+C e^{-ikx}&\hspace{1cm} x< 0\\
D e^{ikx}&\hspace{1cm} 0<x,
\end{array}\right.
\eea
with
\bea\label{CD}
D=\left(1+\frac{iz}{2k}\right)^{-1}, \hspace{1cm} C=-\frac{iz}{2k} D\ .
\eea
Writing $z=2\lambda(1+i\ve)$ and $k= \lambda q$, it is easy to show that
\bea
|C|^2+|D|^2=\left(1-\frac{2\ve q}{1+\ve^2+q^2}\right)^{-1},
\eea
which is greater than or less than 1 depending on the sign of $\ve$.

\section{Quasi-Hermitian Analysis of the Single Delta Function}

The positive-definite Hermitian metric operator $\eta\equiv e^{-Q}$ is defined by
\bea\label{psH}
H^\dag=\eta H \eta^{-1}.
\eea
We are then instructed to calculate matrix elements by including $\eta$. Thus, for example, the expectation
value of the quantity represented by the operator $A$ in the state $|\psi\rangle$ is
\bea
\langle A \rangle_\psi = \langle\psi|\eta A|\psi\rangle
\eea
assuming that $\psi$ is normalized: $\langle\psi|\eta |\psi\rangle=1$.
An operator $A$ is an observable, with real eigenvalues, if it, like the Hamiltonian, is quasi-Hermitian,
satisfying $A^\dag=\eta A \eta^{-1}$.

Although the metric operator has been worked out for the complex square-well potential by Mostafazadeh\cite{AM-SqWell}
the calculation of the corrected transmission and reflection coefficients is not particularly transparent, because of
the complicated nature of the wave function. Instead we will analyze the single complex $\delta$ function, which,
although not $PT$-symmetric, still has a completely real spectrum and therefore falls within the framework of quasi-Hermiticity.
In this case $\eta$ has
again been calculated by Mostafazadeh\cite{AM-1DF} up to third order in the imaginary
part of the coefficient of the $\delta$ function.

In that reference $\eta$ is given as a perturbation
series in $\ve$,
\bea
\eta=\sum_{r=0}^\infty \ve^r\eta^{(r)},
\eea
 up to order $\ve^3$. The  matrix elements of $\eta^{(0)}$ and $\eta^{(1)}$ are
\bea
\eta^{(0)}(x,y)&=&\delta(x-y)\\
\eta^{(1)}(x,y)&=&\half i\lambda[\theta(xy)e^{-\lambda|x-y|}+ \theta(-xy)e^{-\lambda|x+y|}]\ {\rm sgn}(y^2-x^2)
\eea

If one is to work with this metric, $x$ is no longer an observable. Instead the observable for position is $X$, defined by
\bea
X=\rho^{-1} x \rho,
\eea
where $\rho=\sqrt{\eta}=e^{-Q/2}$.
Thus, indeed, $X^\dag=\rho x \rho^{-1}= \rho (\rho X \rho^{-1}) \rho = \eta X \eta^{-1}$, as required.

The most important property of $\rho$ is that it is not local (note in particular the term in
$\exp[-\lambda|x+y|]$). Its effects are felt not only in the vicinity
of the delta function, but at large distances. Indeed this must be the case if unitarity is to be restored
within the new framework.

Of course it has been known for some time that in many cases the metric operator is non-local,
such as for the potential $ix^3$ \cite{HFJ-ix3, AM-ix3}, and the infinite $PT$-symmetric square-well
potential\cite{MB, CMB-InfSqWell}. However, in such cases the particle is confined, with an exponentially
decreasing wave function so that the range of the non-locality
is limited. Scattering potentials highlight this feature to its full extent because the wave functions of the energy eigenstates do not have compact support.

Let us analyze the scattering problem for the single complex delta function
in the quasi-Hermitian framework. According to Mostafazadeh\cite{AM-1DF,MB} the relevant wave-function
is not $\psi(x)\equiv\langle x|\psi\rangle$, but
\bea\label{Psi}
\Psi(x)\equiv \langle x|\Psi\rangle=\langle x|\rho|\psi\rangle
\eea
According to this definition, the quasi-Hermitian expectation value of the observable $X$ is
\bea
\langle \hat{X}\rangle_\psi&=&\langle\psi|\eta\hat{X}|\psi\rangle\nonumber\\
&=&\langle\Psi|\rho^{-1}\eta\ (\rho^{-1}\hat{x}\rho)\rho^{-1}|\Psi\rangle\\
&=&\langle\Psi|\hat{x}|\Psi\rangle = \int x|\Psi(x)|^2 dx\nonumber
\eea
The probability density is then taken as $\varrho\equiv |\Psi(x)|^2$, and the total probability is
conserved\cite{MB} in time: $d(\int\varrho\  dx)/dt=0$. However, because of the non-locality,
there does not seem to be a corresponding local conservation equation of the
form $\partial \varrho/\partial t + dj/dx=0$. Instead one can show that
\bea
i\frac{\partial\varrho(x)}{\partial t}=\int dy  \left[\Psi^*(x) h(x,y)\Psi(y) - \Psi^*(y) h(y,x)\Psi(x)\right],
\eea
where $h$ is the equivalent Hermitian Hamiltonian $h=\rho H \rho^{-1}$.

According to Eq.~(\ref{Psi}), we have to calculate
\bea
\Psi(x)=\int dy \rho(x,y) \psi(y),
\eea
where, to first order,  $\rho=\eta^{(0)}+\half\ve\eta^{(1)}$, and the original wave-function
$\psi(x)$ is given in Eqs.~(\ref{psi}) and (\ref{CD}).

The result of this calculation is that for $x>0$, and to $O(\ve)$,
\bea\label{Psi>}
\Psi_>(x)=D e^{ikx}+\frac{\ve\lambda k}{2(\lambda^2+k^2)}\big( e^{-ikx}-(C+D)e^{ikx}\big)+O(e^{-\lambda x})
\eea
Since we are concerned with asymptotic states, we will neglect the last term.
One extremely important feature of this equation is that $\Psi_>(x)$ no longer represents a pure outgoing wave
$\propto e^{ikx}$, but now contains an $O(\ve)$ component of an incoming wave as well. Thus the physical
picture of the scattering is completely changed.

However, in the calculation of probabilities to $O(\ve)$ we can neglect this term, since it gives rise
to a contribution of $O(\ve^2)$. Doing so, and using the simplified notation $k= q\lambda$ introduced earlier,
Eq.~(\ref{Psi>}) reduces to
\bea
\Psi_>(x)=e^{ikx}\frac{q}{q+i}\left(1+\frac{\ve }{2(q+i)}\right),
\eea
so that
\bea\label{psi>}
|\Psi_>(x)|^2=\frac{q^2}{q^2+1}\left(1+\frac{\ve q }{q^2+1}\right),
\eea

The corresponding calculation for $x<0$, again ignoring exponentially damped terms, this time
 $\propto e^{\lambda x}$, gives
\bea\label{Psi<}
\Psi_<(x)&=&e^{ikx}\left(1+\frac{\ve\lambda k}{2(\lambda^2+k^2)}\right)
+e^{-ikx}\left(C-\frac{\ve\lambda k}{2(\lambda^2+k^2)}(C+D)\right)\nonumber\\
&&\\
&=&e^{ikx}\left(1+\frac{\ve q}{2(q^2+1)}\right)-\frac{i e^{-ikx}}{q+i}\left(1+\frac{\ve}{2(q+i)}-i\ve\right),\nonumber
\eea
so that
\bea\label{psi<}
|\Psi_<(x)|^2=\left(1+\frac{\ve q}{q^2+1}\right) + \frac{1}{q^2+1}\left(1+\frac{\ve q}{q^2+1}\right)+ \mbox{interference term},
\eea
in which the first term represents the incoming flux and the second term the reflected flux.
Thus we see that each Hermitian flux, for the real delta function $2\lambda\delta(x)$, is multiplied by the common
factor $1+\ve /(q^2+1)$, so that the newly-defined probability is conserved to this order.

\section{Discussion}

Eqs.~(\ref{Psi>}) and (\ref{Psi<}) provide the resolution of the lack of conservation
of probability in the normal quantum mechanical framework, which is achieved by recognizing that
the observable $X$ differs from the usual position variable $x$, even at infinity. Recall that
in the calculation to $O(\ve)$ we were able to ignore the term $\propto e^{-ikx}$ in $\Psi(x)$. However,
it would have to be taken into account in higher orders, and represents a completely different scattering
situation from the original one, where the wave-function $\psi(x)$ represented a wave incoming from the left
and being either reflected or transmitted. Instead, the corresponding $\Psi(x)$ represents a wave with a
component also incoming from the right.

However, the scattering problem should be posed in terms of $\Psi(x)$ instead of $\psi(x)$. Unfortunately
this can only be done in a perturbative fashion. That is, we have to start with $x$ because that is how the
potential is defined.

The issues we have raised are generic, and not due to the fact that the specific potentials we have considered
involve $\delta$ functions or sharp discontinuities in their derivatives. Thus if we consider a general $PT$-symmetric
potential $V(x)=V^*(-x)$, the integrated continuity equation that results is
\bea
\left[\frac{d j}{dx}\right]_{-\infty}^\infty &=& -2\int_{-\infty}^\infty \mbox{Im} V(x) |\psi(x)|^2\nonumber\\
&=&\int_0^\infty \mbox{Im} V(x) (|\psi(x)|^2-|\psi(-x)|^2),
\eea
where $j\equiv -i(\psi^* d\psi/dx - \psi d\psi^*/dx)$ is the standard probability current.
Barring exceptional cancellations, the right-hand side of this equation vanishes only when $\psi^*(-x)=\psi(x)$,
i.e. when the wave-function is itself $PT$-symmetric. However, in the standard scattering set-up whereby a wave enters
from the left and is either reflected or transmitted $\psi$ is definitely not $PT$-symmetric, and $j$ is not conserved.
Thus any quasi-Hermitian description must necessarily involve a redefinition of $x$ at infinity.

The reason the effect of the non-Hermitian potential is felt at large distances is that the plane wave functions
are completely spread out in position, and indeed physically interact with the potential. However, there would
be a problem if the change in the metric caused by the introduction of the potential were to appreciably change
the properties of particles bound by other localized potentials with support far away from it.

As an example, consider the $PT$-symmetric potential
\be
V(x)=-2\alpha\delta(x)+i\lambda(\delta(x-L)-\delta(x+L)),
\ee
made up of a Hermitian binding potential and two non-Hermitian scattering potentials at a distance $L$, which we will
ultimately take to be large. For $\lambda=0$ the system has a bound state with wave-function
$\psi(x)\propto e^{-\alpha|x|}$ and energy $\alpha^2$. While the scattering properties of this potential
will undoubtedly be affected by the presence of the distant non-Hermitian potentials, we would not expect
the bound-state energy or wave-function to be appreciably changed.

Writing the wave function as
\bea\label{WF}
\psi=\left\{\begin{array}{ll}
A e^{\kappa x} &\hspace{1cm} x< -L\\
B e^{\kappa x}+C e^{-\kappa x}&\hspace{1cm} -L < x< 0\\
D e^{-\kappa x}+E e^{\kappa x}&\hspace{1cm} 0< x< L\\
F e^{-\kappa x}&\hspace{1cm} L<x
\end{array}\right.,
\eea
we find
\bea
B&=& A\left(1-\frac{i\lambda}{2\kappa}\right)\nonumber\\
C&=& A\ \frac{i\lambda}{2\kappa}\ e^{-2\kappa L}
\eea
and similarly
\bea\label{DE}
D&=& F\left(1+\frac{i\lambda}{2\kappa}\right)\nonumber\\
E&=& - F\  \frac{i\lambda}{2\kappa}\ e^{-2\kappa L}
\eea
Finally, applying the continuity conditions at $x=0$ we obtain the
eigenvalue equation for $\kappa$, which is most conveniently written as
\bea
\alpha=\kappa+\frac{e^{-2\kappa L}}{1+\left(2\kappa/\lambda\right)^2}
\left[2\alpha+\left(\kappa+\alpha\right)e^{-2\kappa L}\right].
\eea
For large $L$ this reduces to
\bea
\frac{\kappa}{\alpha}\approx 1 -\frac{2 e^{-2\alpha L}}{1+\left(2\alpha/\lambda\right)^2},
\eea
showing that indeed the change in the energy is exponentially small.

 From Eqs.~(\ref{WF}) and (\ref{DE})
we see that although the wave function contains an exponentially growing term in $0<x<L$, its coefficient $E$ is suppressed
relative to $D$ by a factor $e^{-2\kappa L}$. Thus it only becomes comparable with the leading term
$De^{-\kappa x}$ at distances of the order of $L$, when both terms are exponentially small.
The wave-function is in fact $PT$-symmetric, with $\psi(-x)=e^{i\vf}\psi^{*}(x)$, where $\vf$ is an arbitrary
phase. In particular $|A|=|F|$, a condition equivalent to the eigenvalue equation.

We have not attempted to calculate $\eta$ for this example. It is, of course, irrelevant to the calculation
of the energy, which is invariant under the similarity transformation induced by $\eta$. It would undoubtedly
have the same feature of non-locality as that of the previous section, and would necessarily induce a large
change between $\psi(x)$ and $\Psi(x)$ for the scattering problem, where the standard probability current
is not conserved. However, as anticipated, the corresponding changes in the bound-state wave-function would
be exponentially small for large $L$ because they would involve values of $\psi$ at large distances from the
origin.

A plausible generalization from this example is that the treatment of bound-state problems is essentially
unaffected by the introduction of distant non-Hermitian scattering potentials, and that it is only when the system
physically interacts with those potentials that a drastic change in the formalism is
required, if indeed the non-Hermitian potentials are regarded as fundamental rather than effective. In principle
this change
provides a consistent picture of the scattering, although in practice it seems extremely difficult
to carry out, since $\eta$ can typically  be calculated only up to some low order in perturbation theory.\\

\noindent{\bf Acknowledgements}

\noindent I am grateful to Prof.~R.~J.~Rivers for useful discussions.

\end{document}